\documentclass[showpacs,preprintnumbers,amsmath,amssymb,APSl,prd,nofootinbib,superscriptaddress]{revtex4}
\usepackage{amsmath}
\usepackage{amssymb}
\usepackage{amsfonts}
\usepackage{graphicx,bm}
\usepackage{dcolumn}
\usepackage[colorlinks=true]{hyperref}
\usepackage{color,amsxtra}
\usepackage{epsf}
\usepackage{enumerate}
\usepackage{hhline}
\usepackage{array}
\usepackage{tabularx}
\usepackage{subfigure}
\usepackage{fancyhdr}
\usepackage{mathrsfs}

\newcommand{\be}{\begin{equation}}
\newcommand{\ee}{\end{equation}}
\newcommand{\bea}{\begin{eqnarray}}
\newcommand{\eea}{\end{eqnarray}}
\newcommand{\beaa}{\begin{eqnarray*}}
\newcommand{\eeaa}{\end{eqnarray*}}

\newcommand{\nn}{\nonumber \\}

\def\be{\begin{equation}}
\def\ee{\end{equation}}
\def\bea{\begin{eqnarray}}
\def\eea{\end{eqnarray}}

\begin{document}
\title{Analyzing the $H_0$ tension in $F(R)$ gravity models}
%
%
\author{Sergei D. Odintsov}
\email{odintsov@ice.csic.es} \affiliation{Institut de Ci\`{e}ncies de l'Espai,
ICE/CSIC-IEEC, Campus UAB, Carrer de Can Magrans s/n, 08193 Bellaterra (Barcelona),
Spain}
 \affiliation{Instituci\'o Catalana de Recerca i Estudis Avan\c{c}ats (ICREA),
Passeig Luis Companys, 23, 08010 Barcelona, Spain}
\author{Diego S\'aez-Chill\'on G\'omez}
\email{diego.saez@uva.es} \affiliation{Department of Theoretical, Atomic and Optical
Physics, Campus Miguel Delibes, \\ University of Valladolid UVA, Paseo Bel\'en, 7, 47011
Valladolid, Spain}
\author{German~S.~Sharov}
 \email{sharov.gs@tversu.ru}
 \affiliation{Tver state university, Sadovyj per. 35, 170002 Tver, Russia}
 \affiliation{International Laboratory for Theoretical Cosmology,
Tomsk State University of Control Systems and Radioelectronics (TUSUR), 634050 Tomsk,
Russia}

\begin{abstract}
The Hubble constant tension problem is analysed in the framework of a class of modified
gravity, the so-called $F(R)$ gravity. To do so, we explore two models: an exponential
and a power-law $F(R)$ gravities, which includes an early dark energy (EDE) term in the latter.
These models can describe both an early time inflationary epoch and the late time
accelerating expansion of the universe. We confront both models with recent
observational data including the Pantheon Type Ia supernovae sample, the latest
measurements of the Hubble parameter $H(z)$ from differential ages of galaxies (cosmic
chronometers) and separately from baryon acoustic oscillations. Standard rulers data set
from the Cosmic Microwave Background radiation are also included in our analysis. The
estimations of the Hubble constant appear to be essentially depending on the set of
observational data and vary in the range from 68 to 70{.}3 km\,/(s$\cdot$Mpc). The fits
of the other free parameters of the models are also obtained, revealing interesting
consequences.

\end{abstract}
\maketitle

\section{Introduction}\label{Intr}

Among current problems in modern cosmology, the tension among estimations of the Hubble
constant $H_0$ is one of the most striking and irritating for researchers.  Over the
last years such discrepancies among  $H_0$ measurements have been revealed through two
different methods: by the Planck collaboration after collecting and analyzing data from
the cosmic microwave background radiation (CMB) over the last 7 years
\cite{Planck13,Planck15,Planck18}, which provides an estimation of $H_0=67.4\pm0.5$
km\,s${}^{-1}$Mpc${}^{-1}$ (Planck18), and on the other hand by the SH0ES
group of Hubble Space Telescope (HST) \cite{HST18,HST19} with the last estimate given by
$H_0=74.03 \pm1.42$ km\,s${}^{-1}$Mpc${}^{-1}$ (SH0ES19). The HST method includes
measurements of the local distance ladder by combining photometry from Cepheids (and
their period luminosity relation) with other local distance anchors, Milky Way
parallaxes and calibration distances to  Cepheids in the nearest galaxies which are
hosts of Type Ia Supernovae (SNe Ia). In particular, the above estimation for the Hubble
constant by the HST group includes observations of
70 Cepheids in the Large Magellanic Cloud \cite{HST19}.\\

Currently the mismatch among $H_0$ estimations by Planck \cite{Planck18} and HST
\cite{HST19} collaborations exceeds $4\sigma$, as this tension has grown over the last
years, as shown in Table \ref{TH0}. This problem may be dealt as the discrepancy between
observations at early and late cosmological time of our Universe \cite{VerdeTR19}, since
HST group works with late time data while Planck collaboration combines observations from
redshifts in a wide range $0<z<1100$ and uses the standard $\Lambda$CDM model as fiducial model, but the
issue may be approached through a theoretical way. For the former, some researchers have
suggested some different ways for solving the $H_0$ tension problem. Several groups have
analysed the estimations of $H_0$ by using several approaches independent of the Cepheid
distance scale and CMB anisotropies (for a review see
\cite{VerdeTR19,DiValentIntertwined20}). Among these methods with new observational
results for $H_0$, the following approaches may be highlighted: the tip of the red giant
branch (TRGB) method used by the Carnegie-Chicago Hubble Program (CCHP)
\cite{FreedmanTRGB19,FreedmanTRGB20}, lensing objects with strong time delays between
multiple images (H0LiCOW project and others) \cite{Wonglens19,Weilens20,Denzellens20},
CMB-lensing data \cite{BaxterCMBlens20}, maser (megamaser) hosting galaxies
\cite{PesceMaser20}, oxygen-rich variable stars (Miras) \cite{HuangMiras19}.
Some other researchers tried to explain the tension by assuming that Planck or HST
measurements might suffer from systematic errors  \cite{Tens_sys}, but these analysis did not led to convincing solutions of the problem. \\

As shown in Table \ref{TH0}, one can note that most of $H_0$ estimations lie on the
range among  Planck18 and SH0ES19 values, while the local (late-time) $H_0$ measurements
are close to the SH0ES19 value, exceed the early-Universe estimations. Only the CCHP
estimation obtained by TRGB method violates the latter tendency, which have led to some
discussions \cite{VerdeTR19,FreedmanTRGB20}. By comparing these facts, many cosmologists
over the recent years have considered the $H_0$ tension as a hint for new physics
beyond the standard $\Lambda$CDM model with different phenomena in early and/or late
times of the universe evolution \cite{VarDE}\,--\,\cite{WangF20} (see also the extended
list of literature in Ref.~\cite{DiValentIntertwined20}). These analysis suggest several
ways for solving the problem, which can be generally summarized as a mechanism that
shifts the effective $H_0$ value from early to late time Universe under different
factors. The best fit $H_0$ appears to be essentially depending on the mentioned
factors. Following this idea, some scenarios have been studied:
\begin{itemize}
\item Dark energy models with a varying equation of state (EoS), via a varying EoS parameter $w$ or via the dark energy density $\rho_{DE}$ \cite{VarDE}.

\item Scenarios with an early dark energy (EDE) component, reproduced in different frameworks (scalar fields, axions), which becomes important before the epoch of matter-radiation equality $z \simeq 3000$ and then decays after faster than radiation \cite{EDE,EDE1}.

\item Models with evolving or decaying  dark matter into dark radiation or other species \cite{VarDM}.

\item Interacting dark energy and dark matter models \cite{IntDE}.

\item Models with extra relativistic species which can interact or modify the effective number $N_{\mbox{\scriptsize eff}}$ at the recombination era \cite{VarNeff}.

\item  Modified gravity models that emerge at an intermediate epoch, as scalar-tensor theories, $F(R)$ gravity, $F(T)$ and others \cite{MG,WangF20,Zumalacarregui:2020cjh}.
\end{itemize}

\begin{table}[bh]
\centering
 {\begin{tabular}{|l|l|c|l|l|l|}\hline
 Project & Year &  $H_0$ (km\,s${}^{-1}$Mpc${}^{-1}$) & Method &  Refs \\ \hline\hline
 Planck & 2018 & $67.4\pm0.5$   & CMB power spectra+lensing & \cite{Planck18} \\
\hline
 SH0ES (HST)   & 2019 & $74.03 \pm1.42$ & Cepheid distance ladder & \cite{HST19} \\
\hline
 CCHP & 2020 & $69.6 \pm 0.8 \pm 1.7$   & TRGB & \cite{FreedmanTRGB20} \\
\hline
 H0LiCOW & 2019 & $73.3^{+1.7}_{-1.8}$ & 6 strong lenses \& $\Lambda$CDM& \cite{Wonglens19}\\ \hline
 - & 2020 & $75.3^{+3.0}_{-2.9}$   & 7 strong lenses + SNe Ia & \cite{Weilens20}\\ \hline
  - & 2020 & $71.8^{+3.8}_{-3.3}$   & 8 strong lenses \& $\Lambda$CDM& \cite{Denzellens20}\\ \hline
 - & 2020 & $73.5\pm 5.3$   & CMB lensing + SNe Ia  & \cite{BaxterCMBlens20}\\ \hline
 Megamaser & 2020 & $73.9\pm 3.0$ & 6 maser galaxies & \cite{PesceMaser20}\\ \hline
  HST & 2019 & $72.7\pm 4.6$ & 6 Miras in SN Ia host galaxy & \cite{HuangMiras19}\\ \hline
  \end{tabular}
 \caption{Recent estimations of the Hubble constant $H_0$} \label{TH0}}\end{table}

In this sense, modified gravities have been widely studied in the literature in the
framework of cosmology (for a review see \citep{Nojiri:2010wj}). Particularly, $F(R)$
gravity is very well known by the scientific community, with an extensive literature
where many aspects of the theory have been analysed. This modification of GR assumes a
generic function of the Ricci scalar for the gravitational action instead of the linear
term of the Hilbert-Einstein action. Such modification leads to interesting properties
and a rich phenomenology that can solve some of the most important problems in
cosmology, as the origin of dark energy. In this sense, $F(R)$ gravity can reproduce
well the late-time acceleration with no need of additional fields and alleviate the
cosmological constant problem by compensating the large value for the vacuum energy
density predicted by quantum field theories (see \cite{Capozziello:2002rd,
OdintsovOF:2020, HuSawicki07, Nojiri:2007as, delaCruz-Dombriz:2015tye,BambaGL:2010}). In
addition, the same type of modifications of GR have been studied in the framework of
inflation where as in the case of dark energy, $F(R)$ gravities can lead to successful
scenarios that fit perfectly well the constraints on the spectral index of perturbations
as given by the analysis of the CMB (\citep{ Starobinsky:1980te}). With this in mind,
models that unify the dark energy epoch and the inflationary paradigm through the
corrections introduced in the gravitational action have been proposed with a great
success \citep{Nojiri:2005pu,OdintsovSGS:2017,OdintsovSGSFlog:2019, ElizaldeNOSZ11,
NojiriOdOik_Uni:2020,OdintsovOik_UniAx:2019,OdintsovOik_Axion:2020, CognolaENOSZ08}. In
addition, some $F(R)$ gravity models that reproduce late-time acceleration can also
recover GR at local scales where this one is very well tested, leading to the so-called
viable $F(R)$ gravity models \cite{HuSawicki07, Nojiri:2007as}. Hence, $F(R)$ gravity is
particularly of great interest in cosmology.

Hence, supported on the the great knowledge of $F(R)$ gravity and its success on trying
to solve some of the most important problems in cosmology, the possibility of
alleviating the $H_0$ tension problem in this framework may be promising, despite has
not been studied yet exhaustively. Although some efforts are being done, as the analysis
to solve this tension by viable $F(R)$ models, and particularly through the Hu-Sawicki
$F(R)$ model \cite{HuSawicki07} in Ref.~\cite{WangF20}, where concluded that the
Hu-Sawicki gravity cannot reduce the $H_0$ tension. In the present paper, we analyse two
$F(R)$ models and the possibility of alleviating the $H_0$ tension. We confront the
models with observational data and estimate the Hubble constant $H_0$ and other model
parameters by using approaches developed in some previous papers
\cite{OdintsovSGS:2017,OdintsovSGSFlog:2019,OdintsovSGS:2020,Sharov:2016,PanSh:2017,ShSw:2020}.
Here we include in our analysis the following observations: the Type Ia supernovae data
(SNe Ia) from the Pantheon sample survey \cite{Scolnic17}, data connected with cosmic
microwave background radiation (CMB) and extracted from Planck 2018 observations
\cite{Planck18,ChenHuangW2018}, and estimations of the Hubble parameter $H(z)$  for
different redshifts $z$ from two different sources: (a) measured from differential ages of
galaxies (in other words, from cosmic chronometers, these  31 data points are analyzed
separately) and (b) $H(z)$ obtained as observable
effect of baryon acoustic oscillations (BAO). We obtain the best fit parameters and compare to the ones from $\Lambda$CDM. \\

The paper is organized as follows. In section \ref{Models}, we introduce $F(R)$ gravity
and the two models we analyse along the paper. Section \ref{Observ} is devoted to  SNe
Ia, $H(z)$ and CMB  observational data. In section \ref{Results} we analyze the results,
estimations for the Hubble constant $H_0$ and other model parameters. Finally, section
\ref{conclusions} gathers the conclusions of the paper.

\section{$F(R)$ gravity models}
 \label{Models}

We can start by reviewing the basics of what is called $F(R)$ gravity, a generalization of the Einstein-Hilbert action that assumes a more complex Lagrangian in terms of the Ricci scalar $R$:
 \begin{equation}
  S = \frac1{2\kappa^2}\int d^4x \sqrt{-g}\,F(R)  + S^{\mbox{\scriptsize matter}}\ .
 \label{Act1}\end{equation}
 Here $\kappa^2=8\pi G$ and $S^{\mbox{\scriptsize matter}}$ is the matter action. In the present paper, we are interested in analyzing the $H_0$ tension problem in the framework of $F(R)$ gravity with the following general form for the action \cite{NojiriOdOik_Uni:2020}:
 \begin{equation}
 F(R)=R  + F_{\mbox{\scriptsize inf}} + F_{\mbox{\scriptsize EDE}} + F_{\mbox{\scriptsize
  DE}}\ .
  \label{FallEDE}
\end{equation}
The first term here is the Einstein-Hilbert action, $F_{\mbox{\scriptsize inf}} $ is assumed to describe the early-time inflation \cite{NojiriOdOik_Uni:2020,OdintsovOik_UniAx:2019,OdintsovOik_Axion:2020} becoming negligible at late times $z<3000$ (only this  epoch is visible in our observational data), whereas $F_{\mbox{\scriptsize DE}} $ plays the role of dark energy, dominating at late times, and is the object under study in this paper. Finally, we have added an extra term, $F_{\mbox{\scriptsize EDE}}$, that behaves as an early dark
energy (EDE) term, i.e. mimics an effective cosmological constant at intermediate times but then dilutes along the expansion, helping to suppress some inadequate behaviors during the intermediate
phases between the matter-radiation equality and recombination \cite{EDE,EDE1}. The general field equations for $F(R)$ gravity are obtained by varying the action (\ref{Act1}) with respect to the metric $g_{\mu\nu}$, leading to:
$$ F_R R_{\mu\nu}-\frac F2
g_{\mu\nu}+\big(g_{\mu\nu}g^{\alpha\beta}\nabla_\alpha\nabla_\beta-\nabla_\mu\nabla_\nu\big)F_R
=\kappa^2T_{\mu\nu}\ ,
 $$
 where $R_{\mu\nu}$ and $T_{\mu\nu}$ are the Ricci and energy-momentum tensors respectively. By assuming a spatially-flat Friedman-Lema\^itre-Robertson-Walker (FLRW) space-time 
 $$ds^2 =-dt^2 + a^2(t)\,d\mathbf{x}^ 2$$
with the scale factor $a(t)$, the FLRW equations in $F(R)$ gravity are obtained:
 \begin{eqnarray}
 \frac{dH}{d\log a}&=&\frac{R}{6H}-2H,\nn
 \frac{dR}{d\log a}&=&\frac1{F_{RR}}\bigg(\frac{\kappa^2\rho}{3H^2}-F_R+\frac{RF_R-F}{6H^2}\bigg),
 \label{eqR}\\
 \frac{d\rho}{d\log a}&=&-3(\rho+p). \label{cont} \end{eqnarray}
The continuity equation (\ref{cont}) can be easily solved for dust matter $\rho_m$ and radiation $\rho_r$ and yields
  \begin{equation}
 \rho=\rho_m^0a^{-3}+ \rho_r^0a^{-4}=\rho_m^0(a^{-3}+X_r a^{-4})\ .
 \label{rho}\end{equation}
 Here $a=1$, $\rho_m^0$ and $\rho_r^0$ are the present time values of the scale factor and the matter densities, while we assume the following estimation for the ratio among densities as provided by Planck \cite{Planck13}:
  \begin{equation}
  X_r=\frac{\rho_r^0}{\rho_m^0}=2.9656\cdot10^{-4}\ . \label{Xrm}
  \end{equation}

The aim of this paper is to explore and compare two clases of $F(R)$ models of the type  described by (\ref{FallEDE}). In both cases, we neglect the inflationary term given by $F_{inf}$ and assume  some initial conditions that mimic $\Lambda$CDM model at large redshifts, namely \cite{OdintsovSGS:2017}:
\begin{equation}
 \frac{H^2}{(H^{*}_0)^2}=\Omega_m^{*} \big(a^{-3}+ X^{*}a^{-4}\big)+\Omega_\Lambda^{*},\qquad
 \frac{R}{2\Lambda}=2+\frac{\Omega_m^{*}}{2\Omega_\Lambda^{*}}a^{-3}, \qquad
 a\to0.
  \label{asymLCDM}\end{equation}
Here the index $*$ refers to parameters as given in the $\Lambda$CDM model. In
particular, $\Omega_\Lambda^{*}=\frac\Lambda{3(H^{*}_0)^2}$ and $H^{*}_0$ is the
Hubble constant in the $\Lambda$CDM scenario as measured today. However, the late-time evolution for the $F(R)$ models deviates from these initial conditions and consequently from $\Lambda$CDM model, such that the above parameters measured today for our models will be different:
 $$
 H_0\ne H^{*}_0, \qquad \Omega_m^0\ne \Omega_m^{*}\ ,
 $$
Nevertheless, these parameters are connected among themselves \cite{HuSawicki07,OdintsovSGS:2017}:
 \begin{equation}
 \Omega_m^0H_0^2=\Omega_m^{*}(H^{*}_0)^2=\frac{\kappa^2}3\rho_m(t_0),
 \qquad  \Omega_\Lambda H_0^2=\Omega_\Lambda^{*}(H^{*}_0)^2=\frac{\Lambda}3\ .
  \label{H0Omm}\end{equation}
It is also convenient as shown below, to redefine the Hubble parameter and the Ricci scalar as dimensionless functions:
\begin{equation}
E=\frac{H}{H_0^{*}},\qquad  {\cal R}=\frac{R}{2\Lambda}\ .
   \label{ER}\end{equation}
The first model is given by the following exponential function
\cite{CognolaENOSZ08,BambaGL:2010,ElizaldeNOSZ11,OdintsovSGS:2017}:
 \begin{equation}
  F(R)=R   + F_{\mbox{\scriptsize DE}}=R-2\Lambda\bigg[1-\exp\Big(-\beta\frac{R}{2\Lambda}\Big)\bigg]\ .
 \label{Fexp}
\end{equation}
Note that this exponential model turns out $\Lambda$CDM model at the limit $\beta\to\infty$. Moreover, at large redshifts, the model also recovers $\Lambda$CDM as the curvature becomes large enough $R>>\Lambda/\beta$. Hence, physical solutions for this $F(R)$ action tend asymptotically to $\Lambda$CDM solutions at large redshifts, such that the above initial conditions (\ref{asymLCDM}) results convenient for the equations. By using the dimensionless variables defined in (\ref{ER}), the corresponding system of equations (\ref{eqR}) can be rewritten as:
\begin{eqnarray}
\frac{dE}{d\log a}&=&\Omega_\Lambda^{*}\frac{{\cal R}}{E}-2E\ , \nn   
\frac{d{\cal R}}{d\log a}&=&\frac{e^{\beta{\cal
R}}}{\beta^2}\bigg[\Omega_m^{*}\frac{a^{-3}+ X_ra^{-4}} {E^2}-1+\beta e^{-\beta {\cal
R}}+\Omega_\Lambda^{*}\frac{1-(1+\beta {\cal R})\,e^{-\beta {\cal R}}}{E^2}\bigg]\ .
  \label{eqR2}\end{eqnarray}

This system of equations can be solved by integrating over the independent variable $x=\log a=-\log(z+1)$ and assuming the initial conditions (\ref{asymLCDM}) at the point $x_i$, where
$e^{-\beta {\cal R}(x_i)}\in(10^{-9},10^{-7})$ and our model mimics $\Lambda$CDM (for more details see Ref.~\cite{OdintsovSGS:2017}). Then, we confront the model with the observational data by fitting the free parameters and keeping in mind that $H(z)=H_0^*E(z)$ with the true value for the Hubble parameter today being  $H_0=H_0^*E(z=0)$ and also the relation (\ref{H0Omm}) for the matter density. \\

Following the same procedure, a second $F(R)$ model with a power-law of the Ricci scalar is analysed, which is described by the gravitational action \cite{NojiriOdOik_Uni:2020,OdintsovOik_UniAx:2019,OdintsovOik_Axion:2020}:
 \begin{equation}
  F(R)=R -2\Lambda\gamma \bigg(\frac{R}{2\Lambda}\bigg)^\delta + F_{\mbox{\scriptsize
  EDE}},\qquad
  F_{\mbox{\scriptsize EDE}}=-\alpha\cdot2\Lambda
  \frac{R^{m-n}(R-R_0)^n}{R_0^{\ell+m}+R^{\ell+m}}
 \label{FRdel}
\end{equation}
Note that here the so-called early dark energy (EDE)
term $F_{\mbox{\scriptsize EDE}}$ is included where $\alpha$, $\ell$, $m$, $n$, $R_0$ are  constants, and the curvature scale $R_0$ corresponds to the Ricci scalar value for the epoch $1000\le z\le 3000$ (see Ref.~\cite{NojiriOdOik_Uni:2020}). This term can generate a quasi-stable  de Sitter stage at $R=R_0$ as far as $n$ is an odd integer, and $\ell$, $m$ are large enough in absence of matter. Early dark energy is aimed to behave as an effective cosmological constant at the time of recombination which might affect the Hubble parameter measurements from the CMB alleviating the Hubble tension, while the term decays at late times. The EDE model (\ref{FRdel}) can realise such behavior for the appropriate curvature scale $R_0$, such that for $R>>R_0$, the EDE term turns out:
\be
F_{EDE}\sim -\frac{2\alpha\Lambda}{R^\ell}\sim 0\ ,
\ee
which means that becomes irrelevant at the very early universe, while at late times $R<<R_0$, the EDE term leads to:
\be
F_{EDE}\propto \frac{R^{m-n}}{R_0^{m+\ell-n}}\sim 0\ .
\ee
Hence, the corresponding dark energy term in the $F(R)$ function (\ref{FRdel}) dominates over $F_{EDE}$ at late-times.
The EDE term becomes important just before recombination $R\sim R_0$, where plays the role of an effective cosmological constant,
as expected by construction \cite{NojiriOdOik_Uni:2020} and inspired by the EDE terms \cite{EDE1}. On the other hand, in the presence of matter,
the limitations on the parameters $\ell$, $m$, $n$ are connected with the behavior of $F_{RR}$ in Eq.~(\ref{FRdel}). \\

By considering the model (\ref{FRdel}) without the EDE term ($\alpha=0$), the model does
not recover purely the $\Lambda$CDM model at the limit $R\to\infty$. However, this
power-law model may mimic the $\Lambda$CDM asymptotic behavior (\ref{asymLCDM}) at large
curvature $10\le{\cal R}\le10^{10}$, whose solutions are free of divergences and
singularities. In this approach we can numerically solve the system of equations
(\ref{eqR}) by fixing the initial conditions (\ref{asymLCDM}) at large redshifts,
corresponding to $1000\le z \le 3000$. The system of equations (\ref{eqR}) for this case
($\alpha=0$) yields:
 \bea
 \frac{dE}{d\log a}&=&\Omega_\Lambda^{*}\frac{{\cal R}}{E}-2E\ , \nn   
 \frac{d\log{\cal R}}{d\log a}&=&\frac{{\cal R}^{1-\delta}}{\gamma(1-\delta)}
 \bigg[\frac{\Omega_m^{*}(a^{-3}+ X_ra^{-4})+\Omega_\Lambda^{*}\gamma\delta(1-\delta){\cal R}^{\delta}}
 {E^2}-1\bigg]+\frac1{1-\delta}\ .
 \label{eqRde}\eea

\begin{figure}[th]
  \centerline{\includegraphics[scale=0.7,trim=4mm 4mm 5mm 5mm]{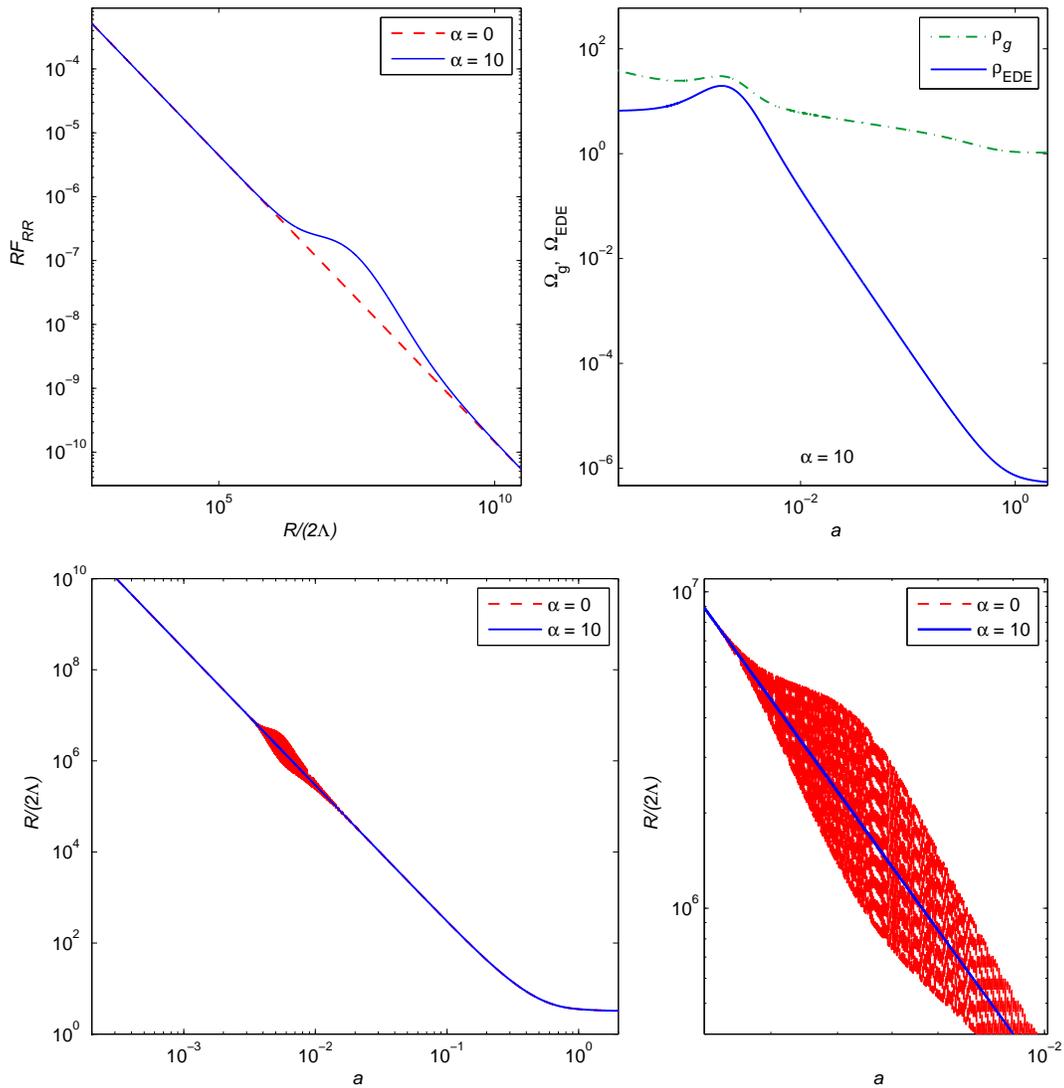}}
\caption{Power-law  model (\ref{FRdel}) with  $\alpha=0$ (red dashed lines) and with
$\alpha=10$ (solid lines): plots for $RF_{RR}(\cal{R})$ (top-left panel) and 
${\cal{R}}(a)$  for  $\gamma=1.5$, $\delta=0.1$, ${\cal{R}}_0=2\cdot10^7$, $m=1$,
$\ell=n=0$, $\Omega_0=0.26$ (bottom panels); in the top-right panel we depict the energy
density evolution for $F(R)$, $\Omega_\mathrm{g}(a)$, and for the EDE term,
$\Omega_\mathrm{EDE}(a)$).}
  \label{F1osc}
\end{figure}

By solving these equations, the corresponding solutions show an undesirable oscillatory
behavior at large $R$ (see Refs.~\cite{NojiriOdOik_Uni:2020,OdintsovOik_Axion:2020}),
especially in the most interesting limit $\delta\ll1$. An example of these oscillations
is depicted in the bottom panels of Fig.~\ref{F1osc}. Such behavior may be controlled
via a choice of the initial conditions (they were optimized in the case shown in the
figure) but can not be completely suppressed in the framework of this model.

Nevertheless, by including the EDE term, these oscillations can be effectively
suppressed, that is with a regular evolution for ${\cal R}$ and $H$.  However, one
should note that for $n\ge1$ and (or) $\ell\ge1$ the second derivative of
$F_{\mathrm{EDE}}(R)$ changes its sign many times close to $R=R_0$, so $F_{RR}(R)$ in
the denominator of Eq.~(\ref{eqR}) can lead to singularities if $\alpha$ is not small
enough. Due to this reason, if the numbers $n$, $m$ or $\ell$ are large, the value for
$\alpha$ should be small. In this case, $F_{\mathrm{EDE}}$ practically does not
influence on describing observational data. Nevertheless, for the case $m=1$,
$\ell=n=0$, the corresponding EDE term becomes:
 \begin{equation}
 F_{\mathrm{EDE}}=-2\Lambda\alpha
 \frac{R}{R_0+R}\ . \label{EDE100}
\end{equation}
And the denominator $F_{RR}(R)$ in Eq.~(\ref{eqR}) behaves well for such a case (see the
top-left panel in Fig.~\ref{F1osc}) and we can use this term with rather large $\alpha$
to suppress oscillations during the epoch when $R\simeq R_0$ and later.

For this choice, figure \ref{F1osc} depicts the evolution of  the Ricci scalar $ {\cal
R}(a) $  and the effective energy density $\rho_\mathrm{g}(a)$ that accounts for the $F(R)$ contribution. The corresponding EDE contribution $\rho_\mathrm{EDE}(a)$ is also depicted in Fig.~\ref{F1osc}. The oscillatory behavior of the Ricci scalar is clearly shown in the bottom-right panel, which blows up the area of oscillations from bottom-left panel.

The effective energy density $\rho_\mathrm{g}$ describes the contribution of $F(R)$
terms which through the FLRW equation (\ref{eqR}) can be written as follows \cite{NojiriOdOik_Uni:2020}:
 \begin{equation}
 \kappa^2\rho_\mathrm{g}=\frac{RF_R-F}{2}+3H^2(1-F_R)-3H\dot F_R\ . \label{rhog}
\end{equation}
The  EDE contribution $\rho_\mathrm{EDE}$ has a similar expression but just with the contribution $F_{\mathrm{EDE}}$. In the top-right panel, the evolution of the normalized energy densities is shown, which as usual are defined as:
 $$
\Omega_\mathrm{g}=\frac{ \kappa^2\rho_\mathrm{g}}{3(H_0^*)^2},\qquad
\Omega_\mathrm{EDE}=\frac{ \kappa^2\rho_\mathrm{EDE}}{3(H_0^*)^2}\, .
 $$

Particularly, for $\Omega_\mathrm{g}$ the equation (\ref{rhog}) yields:
 $$
\Omega_\mathrm{g}(a)=\Omega_\Lambda^{*}\bigg({\cal
R}F_R-\frac{F}{2\Lambda}\bigg)+3E^2\bigg(1-F_R-\frac{dF_R}{d\log a}\bigg)$$
 
Similar expression is obtained for $\Omega_\mathrm{EDE}$ by considering $F_{\mathrm{EDE}}$
instead of $F(R)$. One can see in  Fig.~\ref{F1osc} that $\Omega_\mathrm{EDE}$ peak
bends the $\Omega_\mathrm{g}(a)$ curve, while decays for late-times.\\

 Thus, non-oscillating and non-diverging solutions of the model  (\ref{FRdel}) with the
EDE term (\ref{EDE100}) can be obtained and confronted with the observational data. As
pointed above, here we can see clearly that the EDE term (\ref{EDE100}) behaves as an
effective cosmological constant for $R\geq R_0$ and decays for $R<<R_0$, playing the
role of EDE terms at the time of recombination,
which might provide a way to solve the Hubble tension. \\


\section{Observational data}\label{Observ}

Here we are interested to confront the models described in the previous section in order to obtain the best fit for the free parameters and particularly the best fit for $H_0$ when using different data sources. As mentioned above, these observations include: (a) Type Ia
supernovae (SNe Ia) data from  Pantheon sample \cite{Scolnic17}; (b) estimates of the
Hubble parameter $H(z)$ from cosmic chronometers and line-of-sight BAO  and (c) CMB data
from Planck 2018 \cite{Planck18,ChenHuangW2018}.\\

For the SNe Ia we use the  Pantheon sample database  \cite{Scolnic17} with
$N_{\mbox{\scriptsize SN}}=1048$ points and compare the corresponding SNe Ia distance
moduli  $\mu_i^{obs}$ at redshift $z_i$ from the catalog with their theoretical values by minimizing the $\chi^2$ function:
 \begin{equation}
\chi^2_{\mbox{\scriptsize SN}}(p_1,\dots)=\min\limits_{H_0}
\sum_{i,j=1}^{N_{\mbox{\scriptsize SN}}}
 \Delta\mu_i\big(C_{\mbox{\scriptsize SN}}^{-1}\big)_{ij} \Delta\mu_j,\qquad
 \Delta\mu_i=\mu^{th}(z_i,p_1,\dots)-\mu^{obs}_i\ ,
 \label{chiSN}\end{equation}
Here $C_{\mbox{\scriptsize SN}}$ is the covariance matrix  \cite{Scolnic17} and $p_j$ are the
free model parameters, whereas the distance moduli is given by:
 $$ \mu^{th}(z) = 5
\log_{10} \frac{D_L(z)}{10\mbox{pc}}, \qquad D_L (z)= (1+z)\, D_M,\qquad D_M(z)= c
\int\limits_0^z\frac{d\tilde z}{H(\tilde
 z)}$$
In the expression (\ref{chiSN}) the Hubble constant $H_0$ is considered as a nuisance parameter
{\cite{OdintsovSGS:2017,OdintsovSGSFlog:2019,OdintsovSGS:2020,Sharov:2016,PanSh:2017}
for SNe Ia data, so can be marginalized and estimations can not be obtained for $H_0$ from $\chi^2_{\mbox{\scriptsize SN}}$. However, this provides very important information when fitting the other model parameters. \\

On the other hand, the Hubble parameter data  $H(z)$ are obtained by two different ways of estimation \cite{OdintsovSGS:2017,OdintsovSGSFlog:2019,OdintsovSGS:2020,Sharov:2016,PanSh:2017,ShSw:2020}. The first one is the cosmic chronometers (CC), i.e. estimations of $H(z)$
by using galaxies of different ages $\Delta t$ located closely in terms of the redshift $\Delta z$,
$$ 
 H (z)= \frac{\dot{a}}{a}= -\frac{1}{1+z}
\frac{dz}{dt} \simeq -\frac{1}{1+z} \frac{\Delta z}{\Delta t}.
 $$
Here we consider 31 CC $H(z)$ data points given in Ref.~\cite{HzData}. In the second method $H(z)$ values are estimated from the baryon acoustic oscillation
(BAO) data along the line-of-sight direction. In this paper we use 36 $H_{\mbox{\scriptsize BAO}}(z)$ data points from Refs.~\cite{H_BAO,H_BAO_Zh} that can be found in \citep{SharovVas2018}. For a particular cosmological model with free parameters $p_1, p_2,\dots$, we calculate the $\chi^2$ function by using the CC $H(z)$ data and  the full set CC + $H_{\mbox{\scriptsize BAO}}$ separately, as follows:
function
\begin{equation}
    \chi_H^2(p_1,\dots)=\sum_{j=1}^{N_H}\bigg[\frac{H(z_j,p_1,\dots)-H^{obs}(z_j)}{\sigma _j}  \bigg]^2
    \label{chiH}
\end{equation}
Note that $H_{\mbox{\scriptsize BAO}}$ data points are correlated with BAO angular distances, such that are not considered in other analysis (see Refs.~\cite{OdintsovSGS:2017,OdintsovSGSFlog:2019,OdintsovSGS:2020}). Nevertheless, here we do not use BAO angular distances, such that we avoid any correlation. \\

%

The last source used along the paper is the data from the cosmic microwave background radiation (CMB) that are given by the following observational parameters \cite{ChenHuangW2018}
  \begin{equation}
  \mathbf{x}=\big(R,\ell_A,\omega_b\big),\qquad R=\sqrt{\Omega_m^0}\frac{H_0D_M(z_*)}c,\quad
 \ell_A=\frac{\pi D_M(z_*)}{r_s(z_*)},\quad\omega_b=\Omega_b^0h^2\ ,
 \label{CMB} \end{equation}
where $z_*=1089.80 \pm0.21$ is the photon-decoupling redshift  \cite{Planck18}, while $h=H_0/[100\,\mbox{km}\mbox{s}^{-1}\mbox{Mpc}^{-1}]$, the radiation-matter ratio
$X_r=\Omega_r^0/\Omega_m^0$ is given in(\ref{Xrm}), and we consider the current baryon
fraction $\Omega_b^0$  as the nuisance parameter to marginalize over. The corresponding $\chi^2$ function is:
 \begin{equation}
\chi^2_{\mbox{\scriptsize CMB}}=\min_{\omega_b}\Delta\mathbf{x}\cdot
C_{\mbox{\scriptsize CMB}}^{-1}\big(\Delta\mathbf{x}\big)^{T},\qquad \Delta
\mathbf{x}=\mathbf{x}-\mathbf{x}^{Pl}\ .
 \label{chiCMB} \end{equation}
where ~\cite{ChenHuangW2018}
  \begin{equation}
  \mathbf{x}^{Pl}=\big(R^{Pl},\ell_A^{Pl},\omega_b^{Pl}\big)=\big(1.7428\pm0.0053,\;301.406\pm0.090,\;0.02259\pm0.00017\big)\ ,
   \label{CMBpriors} \end{equation}
with free amplitude for the lensing power spectrum, from Planck collaboration 2018 data \cite{Planck18}. The covariance matrix $C_{\mbox{\scriptsize CMB}}=\|\tilde C_{ij}\sigma_i\sigma_j\|$, the expression $r_s(z_*)$ and other details are well described in
Refs.~\cite{OdintsovSGSFlog:2019,OdintsovSGS:2020}} and \cite{ChenHuangW2018}.

\section{Results and discussion}\label{Results}

\begin{figure}[th]
  \centerline{\includegraphics[scale=0.71,trim=4mm 4mm 5mm 5mm]{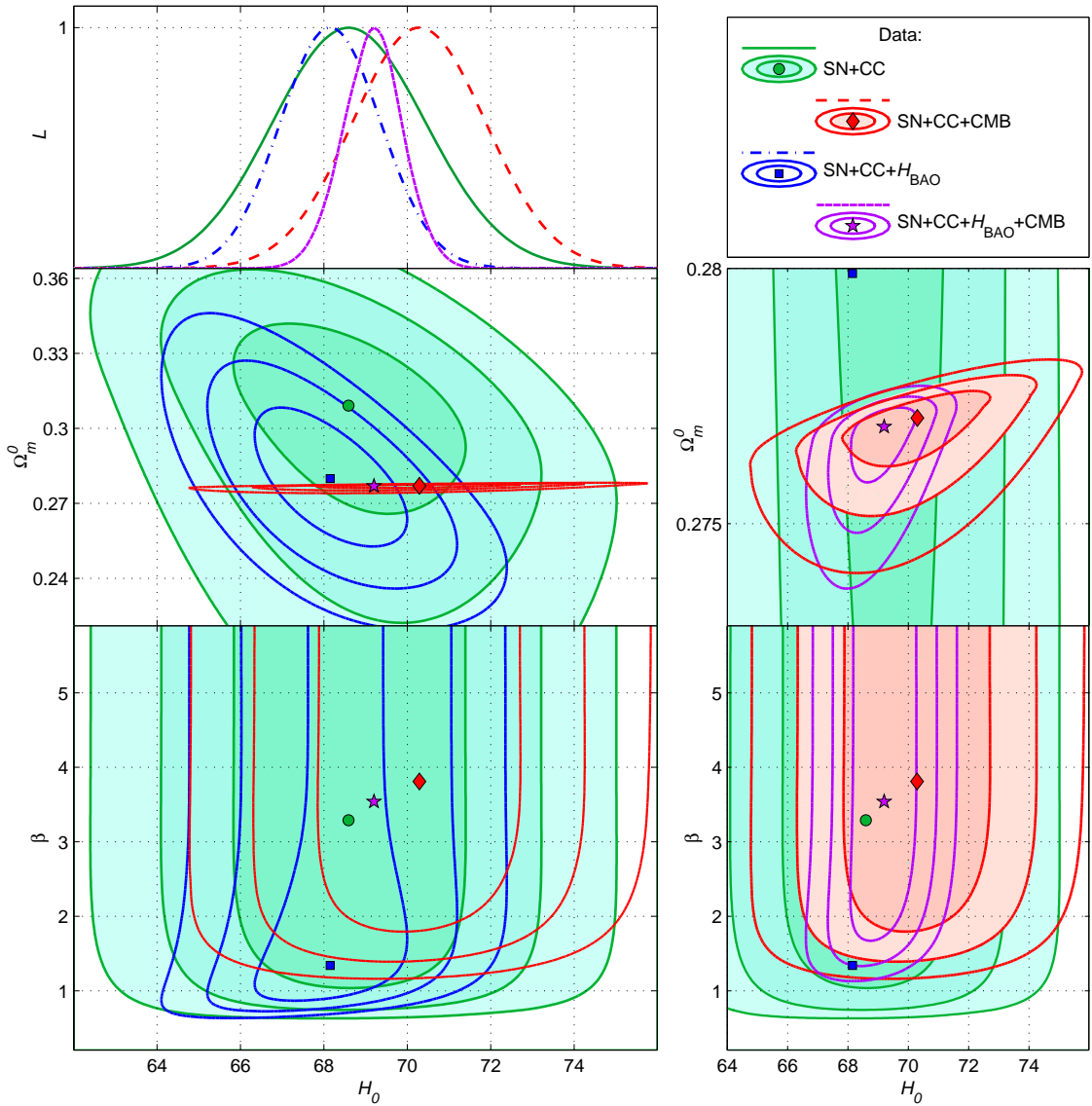}}
\caption{Exponential model (\ref{Fexp}):  $1\sigma$, $2\sigma$ and $3\sigma$ CL  contour
plots for $\chi^2(H_0,\Omega_m^0)$,  $\chi^2(H_0,\beta)$ and  likelihood functions $
{\cal L}_H(H_0)$ for the 4 sets of observational data. }
  \label{F2ex}
\end{figure}

Let us now fit the corresponding models parameters through the $\chi^2$ functions as given in (\ref{chiSN}),(\ref{chiH}) and (\ref{chiCMB}) for each $F(R)$ model. We consider separately
the SNe Ia and $H(z)$ CC (or CC +  $H_{\mbox{\scriptsize BAO}}$) data,
  \begin{equation}
  \chi^2_{\mbox{\scriptsize SN}+H}=\chi^2_{\mbox{\scriptsize SN}}+\chi^2_H
 \label{chiSNH} \end{equation}
and the same SNe Ia and $H(z)$ data with the CMB contribution (\ref{chiCMB})
  \begin{equation}
  \chi^2_{\mbox{\scriptsize SN}+H+{\mbox{\scriptsize CMB}}}=\chi^2_{\mbox{\scriptsize SN}}+\chi^2_H+\chi^2_{\mbox{\scriptsize
  CMB}}\,.
 \label{chitot} \end{equation}
We follow this procedure as the CMB data (\ref{chiCMB}) with narrow  priors (\ref{CMBpriors})
produce the most tight limitations on the model parameters, particularly on
the density matter parameter due to the factor $\sqrt{\Omega_m^0}$ in Eq.~(\ref{CMB}) (see
Fig.~\ref{F2ex}). Thus, for the two $F(R)$ models we analyze four different sets of data:
 \begin{equation} \begin{array}{ll}
  \mbox{SNe Ia + CC},& \mbox{ \ \ SNe Ia + CC + }H_{\mbox{\scriptsize BAO}};\\
 \mbox{SNe Ia + CC + CMB},\quad& \mbox{ \ \ SNe Ia + CC + }
H_{\mbox{\scriptsize BAO}}\mbox{ + CMB} . \end{array}
  \label{4data} \end{equation}
Following the way of maximizing the likelihood, we obtain the corresponding distributions and contour plots for the free parameters for both $F(R)$ models.\\

The exponential model (\ref{Fexp}) owns four free parameters: $H_0$, $\Omega_m^0$,
$\Omega_\Lambda$, $\beta$ or equivalently $H_0^*$, $\Omega_m^*$, $\Omega_\Lambda^*$,
$\beta$. All these parameters are considered with flat priors within their natural limitations (positive values).
This approach does not make problems for our models, because all $\chi^2$ functions for
the sets  (\ref{4data}) have different minimums and grow rather quickly when
values of the parameters are far away from such points and beyond the physical limits. By the
results, we see that $1\sigma$ and even $3\sigma$ confidence level domains lie inside the  physically admissible 
regions of the parameter space.

Fig. \ref{F2ex} shows that the mentioned $1\sigma$ CL domains include also the
limiting points with $\beta\to\infty$ when our exponential model tends to the
$\Lambda$CDM model. This behavior is natural: in the $\Lambda$CDM limit  the  model
(\ref{Fexp}) successfully describes the observational data  (\ref{4data}).

This approach with flat priors for the free parameters was previously realised in Refs.~\cite{OdintsovSGS:2017,OdintsovSGSFlog:2019,OdintsovSGS:2020}. At each point of a
2-parameter plane $p_1-p_2$ (for example, $H_0=p_1$ and $\Omega_m^0=p_2$ in
the case of the $H_0-\Omega_m^0$ plane) we search for the minimum of the $\chi^2$ function over
the two remaining  parameters, testing $\chi^2$ in the $p_3-p_4$ plane in the box with fixed
size but moving center. The position of this center depends on previous calculations.

 One should note that the parameter
 $\Omega_\Lambda$ may be considered as a conditionally free parameter because the
$\chi^2$ functions (\ref{chiSNH}) and  (\ref{chitot}) have sharp minimums along the line
$\Omega_m^0+\Omega_\Lambda\simeq \xi(\beta)$, where $\xi(\beta)\to1$ in the $\Lambda$CDM
limit $\beta\to\infty$. Figure \ref{F2ex} depicts the contours at $1\sigma$ (68.27\%),
$2\sigma$ (95.45\%) and $3\sigma$ (99.73\%) for the four data sets given in
(\ref{4data}) when considering the exponential model (\ref{Fexp}). The planes
$H_0-\Omega_m^0$ and $H_0-\beta$ are obtained by maximizing the likelihood (minimizing
the $\chi^2$) over the other parameters, while the absolute maximums are described by
circles, stars etc. The right panels depict the same contour plots but with additional
details, in particular, the  second one is re-scaled along the $\Omega_m^0$ axis. The
corresponding one-parameter distributions shown in the top left panel corresponds to the
likelihood function for $H_0$ after maximizing over the other parameters:
\begin{equation}
{\cal L}_j(H_0)\sim\exp(-\chi^2_j(H_0)/2)\ .
  \label{likeli}
\end{equation}

\begin{figure}[th]
  \centerline{\includegraphics[scale=0.68,trim=6mm 4mm 5mm 1mm]{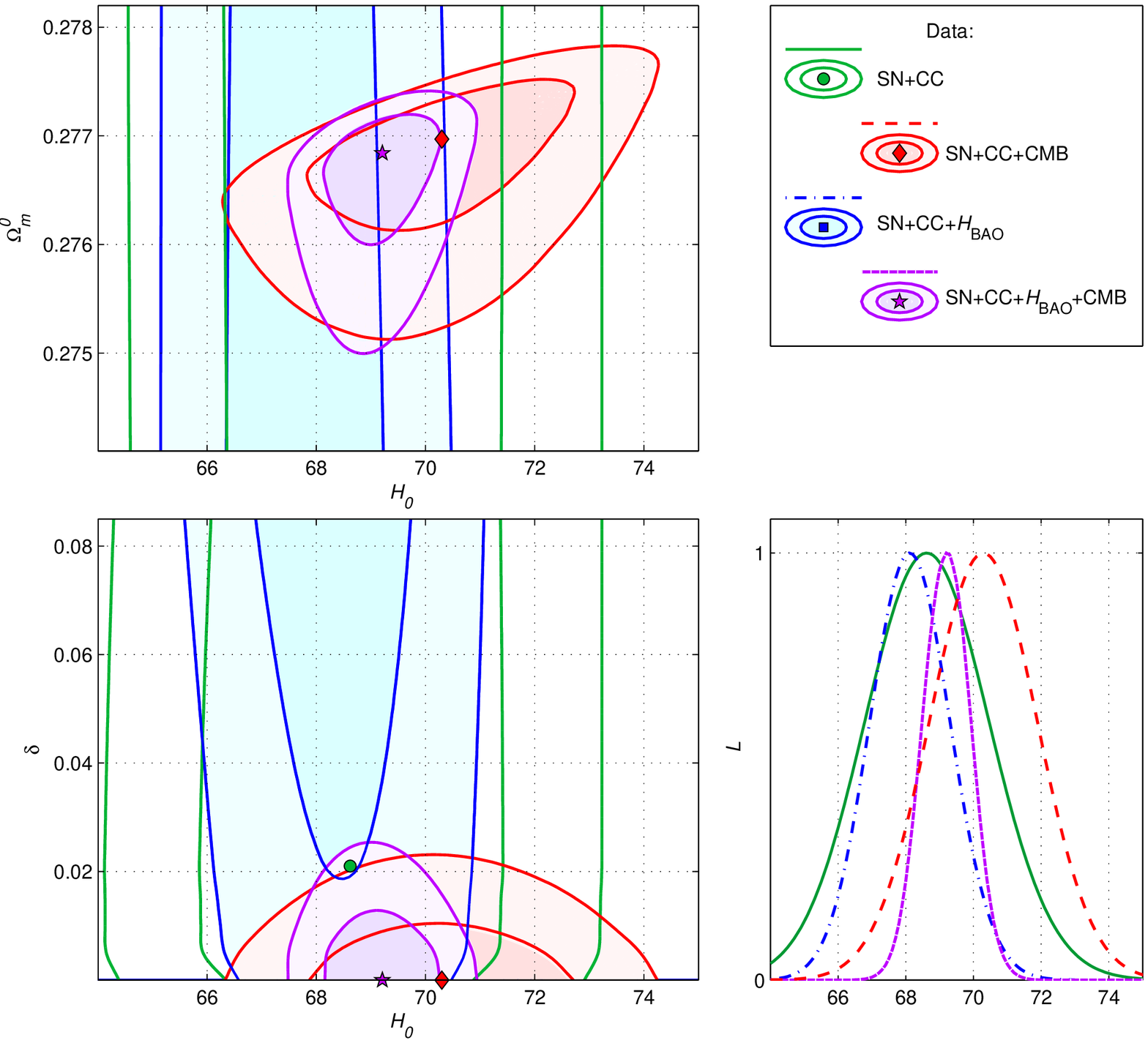}}
\caption{Power-law  model (\ref{FRdel}) with the EDE term:  $1\sigma$ and  $2\sigma$  CL
contour plots in $H_0-\Omega_m^0$, $H_0-\delta$ planes and  likelihood functions $ {\cal
L}_j(H_0)$.}
  \label{F3}
\end{figure}

Following the same procedure, the fits of the free parameters for the power-law  model
(\ref{FRdel}) with the  EDE term (\ref{EDE100}) are obtained by the data sets
(\ref{4data}). This model has the free parameters: $H_0$, $\Omega_m^0$,
$\Omega_\Lambda$, $\gamma$, $\delta$, $\alpha$ and $R_0$. Although the EDE factor
$\alpha$ is denoted by $\alpha^*=\log\alpha$ in Table \ref{Estim}. We fix
$R_0/(2\Lambda)=1.2\cdot10^{7}$ corresponding to the epoch before or near the
recombination and work with the remaining 6 parameters. Here $\Omega_\Lambda$  can be
considered as a conditionally free parameter, because the functions
$\chi^2_j(\Omega_\Lambda,\dots)$ behave like in the previous  exponential model.
The  $1\sigma$ and $2\sigma$ contour plots are shown in Fig.~\ref{F3}. In these  contour plots  and in the  likelihood functions (\ref{likeli}) we also  minimize $\chi^2_j$ over the other parameters. \\

Table \ref{Estim} summarizes the results for both $F(R)$ models together with $\Lambda$CDM model, where the minimum $\chi^2$, the best fits of the free parameters and their errors are shown for the different data sets considered here (\ref{4data}).
We can evaluate the three models in Table~\ref{Estim} from the point of view of information
criteria depending on the  number $N_p$ of the free model parameters. In this sense, the
Akaike information criterion \cite{Akaike} $AIC = \min\chi^2_{tot} +2N_p$ provides a way to compare the goodness of the fits. Then, $\Lambda$CDM model with $N_p=2$ gives a better estimation in comparison with the  exponential  $F(R)$ model with $N_p=4$ and the power-law model with $N_p=6$. On the other hand, the minimum $\chi^2_j$ for the  exponential model (\ref{Fexp}) shows a better fit than $\Lambda$CDM model for the four sets (\ref{4data}), with the largest difference for the SNe Ia + CC + $H_{\mbox{\scriptsize BAO}}$ data, where additionally the 1$\sigma$ region for the $\beta$ parameter does not include $\Lambda$CDM model  (recall that this is recovered for $\beta\rightarrow\infty$). The other three sets show similar fits, including $\Lambda$CDM model within the errors for $\beta$. The value of $H_0$ for the best fit depends on the data set, with its maximum  given by $H_0=70.28\pm1.6$ km\,s${}^{-1}$Mpc${}^{-1}$ (SNe Ia + CC + CMB) and its minimum by $H_0=68.15^{+1.21}_{-1.20}$ km\,s${}^{-1}$Mpc${}^{-1}$ (SNe Ia + CC + $H_{\mbox{\scriptsize BAO}}$).\\

One can see that for the power-law model (\ref{FRdel}), which owns 6 free parameters, the absolute minimum $\min\chi^2_j$ is similar in comparison with the
exponential $F(R)$ and the $\Lambda$CDM models for all data sets (\ref{4data}), with a slightly better fit for the case with for SNe Ia + CC +
$H_{\mbox{\scriptsize BAO}}$ data.  This may be explained as follows: only in this case
the best fit value for the parameter  $\delta=0.17^{+0.11}_{-0.095}$ is large enough
for an effective role of the dark energy $F(R)$ term in this model (\ref{FRdel}). While the other data sets (\ref{4data}), especially for SNe Ia + CC + $H_{\mbox{\scriptsize BAO}}$ data, the best fit for $\delta$ is close to zero, but as far as
$\delta\to 0$  the power-law model (\ref{FRdel}) tends to the $\Lambda$CDM model and the EDE term should be strongly limited. \\

\begin{figure}[h]
  \centerline{\includegraphics[scale=0.7,trim=2mm 4mm 5mm 1mm]{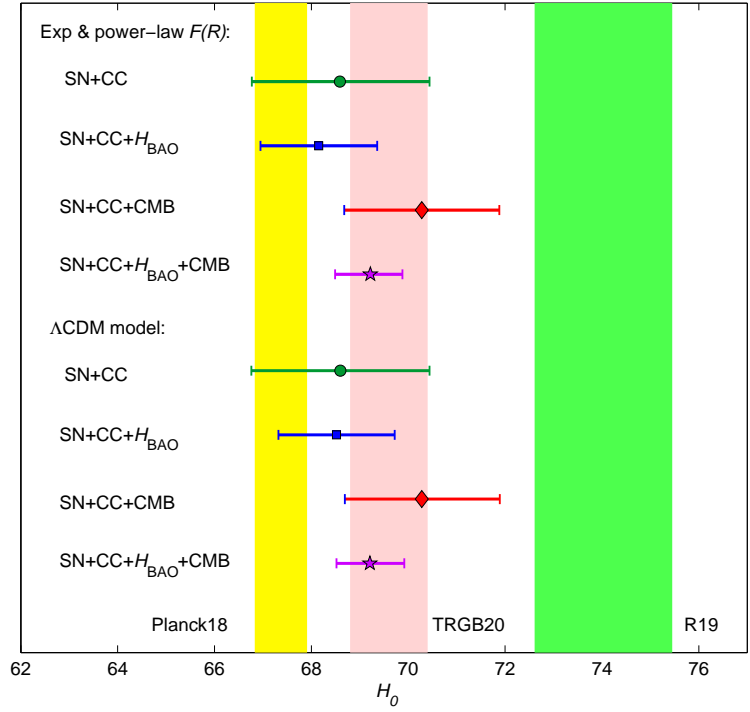}}
\caption{Box plots for the exponential and  power-law $F(R)$ models,together with the $\Lambda$CDM model in comparison with Planck18, TRGB20
and SH0ES (R19) $H_0$ estimations. }
  \label{FWh}
\end{figure}

Concerning the $H_0$ tension problem, Table~\ref{Estim} also gathers the estimations for the Hubble constant for both $F(R)$ models and for $\Lambda$CDM model when confronting the models with the data sets as given in (\ref{4data}). The  best fit values for $H_0$ are very similar for both the exponential model (\ref{Fexp}) as for the power-law model (\ref{FRdel}), and they are depicted together in the box plots of Fig.~\ref{FWh}. Moreover, these estimations are
very close to the $\Lambda$CDM model predictions for three combinations of the observational
data sets (\ref{4data}), whereas for the case SNe Ia + CC + $H_{\mbox{\scriptsize
BAO}}$, the best fit value is shifted to smaller values of $H_0$ for the $F(R)$ models. As shown in Table \ref{Estim}, the case SNe Ia + CC + $H_{\mbox{\scriptsize
BAO}}$ gives the most disparate results in comparison to $\Lambda$CDM model, as the errors on the free parameters does not include $\Lambda$CDM model within the $1\sigma$ region in both $F(R)$ models. The same applies to the $\Omega_m^0$ parameter, which gives almost the same value for all the models when taking three of the data sets, and is slightly different among the $F(R)$ models and $\Lambda$CDM model for the SNe Ia + CC + $H_{\mbox{\scriptsize BAO}}$ set. In addition, the inclusion of CMB data implies much smaller errors on the matter density parameter, as natural due to the factor appearing in (\ref{CMB}). \\

Hence, the results draw an scenario that despite the $H_0$ tension is not alleviated in the $F(R)$ models, it gives an interesting result when considering the SNe Ia + CC + $H_{\mbox{\scriptsize BAO}}$ data set, as excludes $\Lambda$CDM from the 1$\sigma$ region.


\begin{table}[ht]
\begin{tabular}{||l|l|c|c|c|l||}
\hline  Model & Data &  $\min\chi^2/d.o.f$ & $H_0$ & $\Omega_m^0$ & other parameters  \\
\hline
 Expon. &  SN+CC & 1072.78 /1076 & $68.59^{+1.85}_{-1.82}$&  $0.309^{+0.0215}_{-0.0255}$ & $\beta=3.29^{+\infty}_{-1.92}$ \rule{0pt}{1.1em}  \\
\cline{2-6}
 $F(R)$ & SN+CC+$H_{\mbox{\scriptsize BAO}}$& 1081.33 /1112 & $68.15^{+1.21}_{-1.20}$&  $0.2799^{+0.0186}_{-0.018}$ & $\beta=1.34^{+0.99}_{-0.36}$ \rule{0pt}{1.1em}  \\
\cline{2-6}
 & SN+CC+CMB& 1083.70 /1079 & $70.28^{+1.60}_{-1.60}$&  $0.2771^{+0.0004}_{-0.0005}$ & $\beta=3.80^{+\infty}_{-1.62}$ \rule{0pt}{1.1em}  \\
\cline{2-6}
 & SN+all$\,H$+CMB& 1091.88 /1115 & $69.22^{+0.66}_{-0.73}$&  $0.2769^{+0.0003}_{-0.0006}$ & $\beta=3.52^{+\infty}_{-1.54}$ \rule{0pt}{1.1em}  \\
\hline\hline
Power-& SN+CC & 1072.78 /1074 & $68.62^{+1.85}_{-1.83}$&  $0.311^{+0.0295}_{-0.038}$ & $\delta=0.021^{+0.215}_{-0.019}$, $\alpha^*=10.78^{+0.64}_{-1.13}$ \rule{0pt}{1.1em}  \\
\cline{2-6}
law + & SN+CC+$H_{\mbox{\scriptsize BAO}}$& 1080.27 /1110 & $68.20^{+1.14}_{-1.35}$&  $0.2635^{+0.0142}_{-0.0125}$ &  $\delta=0.17^{+0.11}_{-0.095}$, $\alpha^*=10.32^{+1.33}_{-1.27}$  \rule{0pt}{1.1em}  \\
\cline{2-6}
EDE & SN+CC+CMB& 1083.77 /1077 & $70.29^{+1.61}_{-1.60}$&  $0.2770^{+0.0004}_{-0.0005}$ & $\delta=0.001^{+0.005}_{-0.001}$, $\alpha^*\le4.6$  \rule{0pt}{1.1em}  \\
\cline{2-6}
 & SN+all$\,H$+CMB& 1092.02 /1113 & $69.23^{+0.67}_{-0.75}$&  $0.2768^{+0.0003}_{-0.0004}$ & $\delta=0.001^{+0.007}_{-0.001}$, $\alpha^*\le3.9$  \rule{0pt}{1.1em}  \\
\hline\hline
$\Lambda$CDM& SN+CC & 1072.80 /1078 & $68.60^{+1.84}_{-1.84}$&  $0.3095^{+0.0211}_{-0.0205}$  & {\footnotesize$\;\;\Omega_\Lambda=1-\Omega_m^0$}\rule{0pt}{1.1em}  \\
\cline{2-6}
  & SN+CC+$H_{\mbox{\scriptsize BAO}}$ & 1083.06 /1114& $68.52^{+1.20}_{-1.20}$&  $0.2883^{+0.0172}_{-0.0163}$  & \rule{0pt}{1.1em}  \\
\cline{2-6}
  & SN+CC+CMB& 1083.77 /1081 & $70.28^{+1.61}_{-1.59}$&  $0.2771^{+0.0003}_{-0.0003}$ &   \rule{0pt}{1.1em}  \\
\cline{2-6}
 & SN+all$\,H$+CMB& 1092.05 /1117 & $69.21^{+0.71}_{-0.69}$&  $0.2769^{+0.0002}_{-0.0002}$ & \rule{0pt}{1.1em}  \\
\hline
 \end{tabular}
 \caption{The best fit values for $H_0$ (in km\,s${}^{-1}$Mpc${}^{-1}$) and other parameters for the cosmological
scenarios: the exponential $F(R)$ model (\ref{Fexp}) and the power-law model (\ref{FRdel}) in comparison with the $\Lambda$CDM model.}
\label{Estim}
\end{table}

%
%

\section{Conclusions}
\label{conclusions}

In this paper we have explored two $F(R)$ gravity models, where the cosmological
evolution is obtained by solving a dynamical system of equations and then compare to
observational data. The models explored along this paper consist of an exponential
correction to the Hilbert-Einstein action (\ref{Fexp}) and the power-law model
(\ref{FRdel}) with the EDE term (\ref{EDE100}). Keeping in mind their capabilities in
alleviating the  $H_0$ tension between the Planck \cite{Planck18}  and  SH0ES
\cite{HST19} estimations of $H_0$, these models were
confronted with observational data including SNe Ia, 2 types of $H(z)$ estimations and CMB data, and by combining this data
 in four different sets in order to analyze the differences and the possible biased introduced by some of the sets on the parameter estimations.\\

The results are summarized in Table~\ref{Estim}, which includes the best fits for the
Hubble constant $H_0$ and other model parameters with their corresponding $1\sigma$ errors. These  estimations are extracted from the likelihood functions
 ${\cal L}_j(p_i)\sim\exp(-\chi^2_j(p_i)/2)$ of the type (\ref{likeli}) with 1-dimensional normal distributions after marginalising over the rest of parameters. Note that the $1\sigma$ confidence
level intervals for 1D 
 distributions $\chi^2_j(p_i)$ do not coincide in
general with $1\sigma$ domains for 2D 
distributions $\chi^2_j(p_i,p_k)$ in
Figs.~\ref{F2ex} and \ref{F3}. In particular, for the exponential model and the SNe Ia +
CC + $H_{\mbox{\scriptsize BAO}}$ data set the interval $\beta=1.34^{+0.99}_{-0.36}$ is
finite, though the corresponding domain in Fig.~\ref{F2ex} includes the $\beta\to\infty$
limit. This difference appears, because the mentioned $1\sigma$ domains are extracted from
2-dimensional normal distributions ${\cal L}_j(p_i,p_k)$, but the relation
between 1D and 2D normal distributions differs through the corresponding $\chi^2_j$ functions,
connected via minimizing $\chi^2_j(p_i)=\min\limits_{p_j}\chi^2_j(p_i,p_k)$. Remind that
$\chi^2_j(p_i)$ and $\chi^2_j(p_i,p_k)$ are obtained through the minimums of $\chi^2_j(p_i,\dots)$ over all
the other parameters.

 However, for the exponential model and the SNe Ia +
CC + $H_{\mbox{\scriptsize BAO}}$ data we may conclude, that the $\Lambda$CDM
($\beta\to\infty$) limit is not excluded on $1\sigma$ level, if we will base on the 2D
distribution (contour plots) in Fig.~\ref{F2ex}.
 \\

As discussed along the paper, the best fitted values of $H_0$ obtained for each model are very similar, with no particular differences among the
$F(R)$ models. In the box plot depicted in Fig.~\ref{FWh}, the $1\sigma$ error for both
$F(R)$ gravities is the same, and in comparison with the one from the $\Lambda$CDM
model, are also very close for the all data sets (\ref{4data}).
 The  $H_0$ estimations for the two $F(R)$ models are close to the $\Lambda$CDM model.
 Moreover, the best fits for three of the data sets are achieved within the range where the $F(R)$
models tend asymptotically to $\Lambda$CDM, i.e. $\beta\to\infty$  for the exponential
model (\ref{FRdel}) and $\delta\to0$ for the power-law model (\ref{FRdel}).
Nevertheless, for the SNe Ia + CC + $H_{\mbox{\scriptsize BAO}}$ data set, both $F(R)$
models do not include $\Lambda$CDM model within the 1$\sigma$ region, and the best fits
in terms of the minimum value of $\chi^2$ shows a goodness of the fits much better than
$\Lambda$CDM model, and better than the other data
sets. \\

In addition, as shown in Fig.~\ref{FWh}, where the vertical bands refer to $H_0$ estimations from Planck 2018 release \cite{Planck18}, SH0ES (HST) group \cite{HST19} and the intermediate recent value by CCHP group with the red giant branch (TRGB) method \cite{FreedmanTRGB20}, the $H_0$ tension does not show a better behavior within these $F(R)$ models but the tension problem remains. In particular, while the data sets that excludes CMB data fit well the estimations for $H_0$ from Planck 2018 and TRGB, it does not when CMB rulers are included. The same applies to the $\Lambda$CDM model, as shown in Fig.~\ref{FWh}. A further analysis including the whole CMB data is expected to provide similar results, even for the model including an EDE term, as the corresponding fits to the CMB data considered here does not change significantly the Hubble parameter value. \\

Hence, we may conclude that the $F(R)$ models considered in the paper, described by the gravitational actions given in (\ref{Fexp}) and (\ref{FRdel}), can not exhaustively explain the tension between the Planck
\cite{Planck18} and SH0ES  \cite{HST19}  $H_0$ estimations, but they can  alleviate this
tension to some extent. The most interesting result lies on the analysis of both $F(R)$ models with SNe Ia + CC + $H_{\mbox{\scriptsize BAO}}$ data set, as excludes the $\Lambda$CDM limit from the best fit region, a possible signal of the deviations from $\Lambda$CDM and/or of the issue of the data sets.

\section*{Acknowledgements}
SDO acknowledges Project No. PID2019-104397 GB-I00 from MINECO (Spain)  and PHAROS COST Action (CA16214). DS-CG is funded by the University of Valladolid.

\end{document}